\documentclass[pra,showpacs]{revtex4}

\newcommand{\be}{\begin{eqnarray}}
\newcommand{\ee}{\end{eqnarray}}
\newcommand{\balpha}{\mbox{\boldmath $\alpha$}}

\newcommand{\bsigma}{\mbox{\boldmath $\sigma$}}
\begin{document}

\title{\bf Quasi-exact solvability of Dirac equation\\
with Lorentz scalar potential}

\author{Choon-Lin Ho}
\affiliation{Department of Physics, Tamkang University, Tamsui
25137, Taiwan, R.O.C.}

\date{Nov 15, 2005} %\today}

\begin{abstract}

We consider exact/quasi-exact solvability of Dirac equation with a
Lorentz scalar potential based on factorizability of the equation.
Exactly solvable and $sl(2)$-based quasi-exactly solvable
potentials are discussed separately in Cartesian coordinates for a
pure Lorentz potential depending only on one spatial dimension,
and in spherical coordinates in the presence of a Dirac monopole.

\end{abstract}

\pacs{03.65.-w, 03.65.Pm, 02.30.Zz}

\maketitle

\section{Introduction}

Exact solutions in quantum mechanics are hard to come by.  This is
especially so in the case of relativistic wave equations. In
particular, for the Dirac equation only a few exactly solvable
electromagnetic field configurations such as homogeneous magnetic
fields \cite{rabi}, homogeneous electrostatic fields
\cite{sauter}, constant parallel magnetic fields \cite{lam} etc.
are known.

While exact solvability is desirable, in practice it is not always
possible to determine the whole spectrum.  Recently, in non
relativistic quantum mechanics a new class of potentials which are
intermediate to exactly solvable ones and non solvable ones have
been found . These are called quasi-exactly solvable (QES)
problems for which it is possible to determine algebraically a
part of the spectrum but not the whole spectrum
\cite{TU,Tur,GKO,Tur3,Ush}. Usually a QES problem admits a certain
underlying Lie algebraic symmetry which is responsible for the
quasi-exact solutions. Previous studies in QES problems deal
mainly with the non-relativistic equations, and only rather
recently did discussions of the quasi-exactly solvability of the
Dirac equations appear. Particularly, it had been shown that the
Dirac equation  with Coulomb interaction supplemented by a linear
radial Lorentz scalar potential \cite{BK}, and planar Dirac
equation with Coulomb and homogeneous magnetic fields \cite{Ho}
are QES systems. The Pauli equation and the Dirac equation coupled
minimally to a stationary vector potential were also shown to be
QES \cite{HoRoy1}. More recently, added to the list are the
Dirac-Pauli equation coupled non-minimally to external electric
fields \cite{HoRoy2}, and Dirac oscillator with Coulomb
interaction supplemented by a linear radial Lorentz scalar
potential \cite{BN}.

Interest in the Dirac equation with Lorentz scalar potential was
mainly motivated by the MIT bag model of quark confinement
\cite{MIT-bag,Bhad}.  Unfortunately, QES and exactly solvable
Lorentz scalar potentials in such Dirac equation are rather scanty
in the literature. The system considered in \cite{BK} provides the
first example of a QES system.  Motivated by this work,  in
\cite{Znojil} a reparametrization of the radial
 Dirac equation with a scalar and a Lorentz scalar potential was used
to show the existence of infinitely many potentials admitting two
QES states.  However, explicit construction of the QES potentials
was not given, as the reparametrized equations are difficult to
handle in practice. Later, a subclass of QES potentials admitting
doublets and multiplets were constructed by restricting the forms
of the wave functions \cite{SH}.

In \cite{HoRoy1,HoRoy2} we have developed a procedure to construct
exactly-solvable potentials and $sl(2)$-based QES potentials
admitting any finite number of QES states.  The method relies
mainly on the factorizability, or equivalently, on the
supersymmetric structure of the Hamiltonian of the system. In the
present paper we would like to extend the procedure to the Dirac
equation with Lorentz scalar potential. We will determine possible
forms of the potential allowing factorization of the Dirac
equation.  Once factorization is achieved, the forms of the
exactly solvable and QES potentials can be easily determined.

The organization of the paper is as follows.  In Sect.~II we
outline the procedure presented in \cite{HoRoy1,HoRoy2}. Sect.~III
discusses the exact and quasi-exact solvability of the Dirac
equation in one spatial dimension. Sect.~IV is devoted to the
three and four dimensional Dirac equations in Cartesian
coordinates, with the potential depending only one spatial
dimension. In Sect.~V the Dirac equation with a Lorentz scalar and
a magnetic monopole potential is considered. Sect.~VI concludes
the paper.

\section{Outline of the procedure}

In this section we shall outline the general procedure we adopted
in\cite{HoRoy1,HoRoy2} for determining exact/quasi-exact
solvability of the Dirac and the Pauli equation. This procedure
makes full use of the close connection between quasi-exactly
solvable systems and supersymmetry, or factorizability of the
system.

Suppose for a Dirac system one can reduce the corresponding
multi-component equations to a set of one-variable equations
possessing one-dimensional supersymmetry after separating the
variables in a suitable coordinate system. Typically the set of
equations takes the form \be \left(\frac{d}{dr} +
W(r)\right)\psi_- &=& {\cal E}^+ \psi_+~,\nonumber\\
\left(-\frac{d}{dr} + W(r)\right)\psi_+ &=& {\cal E}^ - \psi_-~,
\label{susy} \ee where $r$ is the basic variable, e.g. the radial
coordinate, and $\psi_\pm$ are, say, the two components of the
radial part of the Dirac wave function. The superpotential $W$ is
related to the external field configuration. ${\cal E}^\pm$ may
involve the energy, the mass of the particle, and perhaps some
conserved quantum numbers. We can rewrite this set of equations as
\be A^-A^+\psi_-&=&\epsilon \psi_-~,\\ A^+A^-\psi_+&=& \epsilon
\psi_+~, \ee with
\be
A^\pm\equiv \pm \frac{d}{dr}  +W~, ~~~\epsilon\equiv {\cal E}^+
{\cal E}^-~. \label{A} \ee Explicitly, the above equations read
\be
\left(-\frac{d^2}{dr^2} + W^2 \mp W^\prime\right)\psi_\mp =
\epsilon \psi_\mp~. \label{susy-1} \ee
 Here and below the prime indicates
differentiation with respect to the basic variable.
Eq.~(\ref{susy-1}) clearly exhibits the supersymmetric structure
of the system. The operators acting on $\psi_\pm$ in
Eq.~(\ref{susy-1}) are said to be factorizable, i.e. as products
of $A^-$ and $A^+$. The ground state, with $\epsilon = 0$,  is
given by one of the following two sets of equations: \be A^+
\psi_-^{(0)}(r)&=& 0~~,~~~ \psi_+^{(0)}(r)=0~;\\ A^-
\psi_+^{(0)}(r)&=& 0~~, ~~~\psi_-^{(0)}(r)=0~, \ee depending on
which solution is normalizable.

One can determine the forms of the external field that admit
 exact solutions of the problem by comparing the forms of the
superpotential $W$ with those listed in Table~4.1 of
\cite{Cooper}.

Similarly, from Turbiner's classification of the $sl(2)$ QES
systems \cite{Tur}, one can determine the forms of $W$, and hence
the forms of external fields admitting QES solutions based on
$sl(2)$ algebra.  The main ideas of the procedure are outlined
below. We shall concentrate only on solution of the upper
component $\psi_-$, which is assumed to have a normalizable zero
energy state.

Eq.~(\ref{susy-1}) shows that $\psi_-$ satisfies the Schr\"odinger
equation $H_- \psi_-=\epsilon \psi_-$, with  \be H_-&=& A^-A^+
\nonumber\\ & =&-\frac{d^2}{dr^2} + V(r) ~,\label{SE} \ee with
\be
V(r)=W(r)^2 -W^\prime(r)~. \label{V} \ee We shall look for $V(r)$
such that the system is QES.  According to the theory of QES
models, one first makes an ``imaginary gauge transformation"
 on the function $\psi_-$
\be
\psi_-(r)= \phi(r) e^{-g(r)}~, \label{f-1} \ee where $g(r)$ is
called the gauge function.  The function $\phi(r)$ satisfies
\be
-\frac{d^2\phi(r)}{dr^2} + 2 g^\prime \frac{d\phi(r)}{dr} +
\left[V(r)+ g^{\prime\prime} - g^{\prime 2}\right]\phi
(r)=\epsilon\phi(r)~. \label{phi} \ee For physical systems which
we are interested in, the phase factor $\exp(-g(r))$ is
responsible for the asymptotic behaviors of the wave function so
as to ensure normalizability. The function $\phi(r)$ satisfies a
Schr\"odinger equation with a gauge transformed Hamiltonian
\be
H_G=-\frac{d^2}{dr^2} + 2W_0(r)\frac{d}{dr}  +\left[V(r)
+W_0^\prime - W_0^2\right]~, \label{HG} \ee where $W_0(r)=g^\prime
(r)$.

Now if $V(r)$ is such that the quantal system is QES, that means
the gauge transformed Hamiltonian $H_G$ can be written as a
quadratic combination of the generators $J^a$ of some Lie algebra
with a finite dimensional representation.  Within this finite
dimensional Hilbert space the Hamiltonian $H_G$ can be
diagonalized, and therefore a finite number of eigenstates are
solvable. For one-dimensional QES systems the most general Lie
algebra is $sl(2)$ .  Hence if Eq.~(\ref{HG}) is QES then it can
be expressed as
\be
H_G=\sum C_{ab}J^a J^b + \sum C_a J^a + {\rm constant}~,
\label{H-g} \ee where $C_{ab},~C_a$ are constant coefficients, and
the $J^a$ are the generators of the Lie algebra $sl(2)$ given by
\be J^+ &=& z^2 \frac{d}{dz} - Nz~,\cr
J^0&=&z\frac{d}{dz}-\frac{N}{2}~,~~~~~~~~N=0,1,2\ldots\cr J^-&=&
\frac{d}{dz}~. \ee
 Here the variables $r$ and $z$ are related by
$z=h(r)$, where $h(\cdot)$ is some (explicit or implicit) function
. The value $j=N/2$ is called the weight of the differential
representation of $sl(2)$ algebra, and $N$ is the degree of the
eigenfunctions $\phi$, which are polynomials in a
$(N+1)$-dimensional Hilbert space with the basis $\langle
1,z,z^2,\ldots,z^N \rangle$ :
\be
\phi=(z-z_1)(z-z_2)\cdots (z-z_N)~. \label{phi-2}
\ee
 The requirement in Eq.~(\ref{H-g}) fixes $V(r)$ and $W_0(r)$, and
$H_G$ will have an algebraic sector with $N+1$ eigenvalues and
eigenfunctions.  For definiteness, we shall denote the potential
$V$ admitting $N+1$ QES states by $V_N$.   From Eqs.~(\ref{f-1})
and (\ref{phi-2}), any one of the $N+1$ functions $\psi_-$ in this
sector has the general form
\be
\psi_-=(z-z_1)(z-z_2)\cdots (z-z_N)\exp\left(-\int^z W_0(r)
dr\right)~, \label{psi-1} \ee where $z_i$ ($i=1,2,\ldots,N$) are
$N$ parameters that can be determined by plugging
Eq.~(\ref{psi-1}) into Eq.~(\ref{phi}).  The algebraic equations
so obtained are called the Bethe ansatz equations corresponding to
the QES problem.

Now comes a crucial observation in our procedure: one can rewrite
Eq.~(\ref{psi-1}) as
\be
\psi_- =\exp\left(-\int^z W_N(r,\{z_i\}) dr\right)~, \label{f2}
\ee with
\be
W_N(r,\{z_i\}) = W_0(r) -  \sum_{i=1}^N
\frac{h^\prime(r)}{h(r)-z_i}~. \label{W} \ee There are $N+1$
possible functions $W_N (r,\{z_i\})$ for the $N+1$ sets of
eigenfunctions $\phi$. Inserting Eq.~(\ref{f2}) into $H_-
\psi_-=\epsilon \psi_-$, one sees that $W_N$ satisfies the Ricatti
equation
\be
W_N^2 - W_N^\prime = V_N - \epsilon_N~, \label{Ricatti} \ee where
$\epsilon_N$ is the energy parameter corresponding to the
eigenfunction $\psi_-$ given in Eq.~(\ref{psi-1}) for a particular
set of $N$ parameters $\{z_i\}$.

From Eqs.~(\ref{SE}), (\ref{V}) and (\ref{Ricatti}) it is clear
how one should proceed to determine the external fields so that
the Dirac equation becomes QES based on $sl(2)$: one needs only to
determine the superpotentials $W(r)$ according to
Eq.~(\ref{Ricatti}) from the QES potentials $V(r)$  classified in
\cite{Tur}. This is easily done by observing that the
superpotential $W_0$ corresponding to $N=0$ is related to the
gauge function $g(r)$ associated with a particular class of QES
potential $V(r)$ by $g^\prime (r)=W_0 (r)$.  This superpotential
gives the field configuration  that allows the weight zero
($j=N=0$) state, i.e. the ground state, to be known in that class.
The more interesting task is to obtain higher weight states (i.e.
$j>0$), which will include excited states.  For weight $j$
($N=2j$) states, this is achieved by forming the superpotential
$W_N(r,\{z_i\})$ according to Eq.~(\ref{W}). Of the $N+1$ possible
sets of solutions of the Bethe ansatz equations, the set of roots
$\{z_1,z_2,\ldots,z_N\}$ to be used in Eq.~(\ref{W}) is chosen to
be the set for which the energy parameter of the corresponding
state is the lowest.

In the following three sections, we will employ this procedure to
classify and construct QES Lorentz scalar potnetials.

\section{Two-dimensional Dirac equation in Cartesian coordinates}

Let us consider a $(1+1)$-dimensional Dirac Hamiltonian
 of the kind
\begin{equation}
H=\alpha p + \beta (m+V_s(x))~, \label{H1}
\end{equation}
where $m$ is the mass of the fermion, $p=-id/dx$, $\alpha$ and
$\beta$ are the Dirac matrices,  and $V_s$ is the Lorentz scalar
potential. Such model is of interest in the theory of nuclear
shell model \cite{Bhad,MSW}, and as a model of the self-compatible
field of a quark system \cite{GMM} . Furthermore, if there exists
zero energy solution, then the theory possesses
  spectral asymmetry and fractional fermion number\cite{Bhad,HoKha}.

It has long been known that the system is supersymmetric
\cite{Cooper1,Nogami}. This can be easily shown as follows. Let
the Dirac matrices be represented in terms of the Pauli matrices
as
\begin{eqnarray}
\alpha_x = \sigma_2~,~~~ \beta = \sigma_1~. \label{rep}
\end{eqnarray}
In this representation the Dirac equation $H\psi =E\psi$ for the
2-component wave function
\begin{eqnarray}
\psi(x)= \left( \begin{array}{c} \psi_- (x)\\ \psi_+ (x)
\end{array}\right)
\end{eqnarray}
takes the form:
\begin{eqnarray}
\left(\frac{d}{dx} +  U(x)\right)\psi_- &=& E\psi_+~,\nonumber\\
\left(-\frac{d}{dx} +  U(x)\right)\psi_+ &=& E\psi_-~~.
\label{susy-1a}
\end{eqnarray}
Here $U(x)\equiv V_s(x) + m$ and $\epsilon=E^2$.
Eq.~(\ref{susy-1a}) is now in the factorized form, with $U(x)$
playing the role of the superpotential. As such, this system can
be dealt with according to the general procedure outlined in the
last section.

\subsection{Exactly solvable cases}

The exactly solvable cases have been classified in Table~4.1 of
\cite{Cooper}. For the present system, there are six types of
exactly solvable field configurations, namely, (i) shifted
oscillator-like; (ii) Morse-potential; (iii) Rosen-Morse II
(hyperbolic); (iv) Scarf II (hyperbolic); (v) Rosen-Morse I
(trigonometric); and (vi) Scarf I (trigonometric).

We note here that the case of linear (shifted oscillator)
potential has been discussed in \cite{linear}, and the Morse case
in \cite{Castro}.

\subsection{Quasi-exactly solvable cases}

In this case, seven classes of QES potential $U(x)$ can be
constructed. These classes correspond to Class I to Class VI, and
Class X in Turbiner's classification \cite{Tur}.  We shall
illustrate the construction of Class I potentials below.
Potentials in the other classes can be constructed accordingly.

The QES potential belonging to Class I has the form
\begin{eqnarray}
V_N(x)=a^2 e^{-2\alpha x}- a\left[\alpha(2N+1)+2b\right]e^{-\alpha
x}+ c\left(2b-\alpha\right)e^{\alpha x} + c^2 e^{2\alpha
x}+b^2-2ac~. \label{V-I}
\end{eqnarray}
Without loss of generality, we assume $\alpha, a,c>0$ for
definiteness.  The corresponding gauge function $g(x)$ is given by
\begin{eqnarray}
g(x)= \frac{a}{\alpha} e^{-\alpha x} + \frac{c}{\alpha} e^{\alpha
x} + b x~. \label{g-I}
\end{eqnarray}
One should always keep in mind that the parameters selected must
ensure convergence of the function $\exp(-g(x))$ in order to
guarantee normalizability of the wave function .  The potential
$V(x)$ that gives the ground state, with energy parameter
$\epsilon\equiv  E^2=0$, is generated by
\begin{eqnarray}
V_0 = U_0^2-U_0^\prime~,
\end{eqnarray}
with
\begin{eqnarray}
 U_0(x)=g^\prime(x)=-a e^{-\alpha x}+ c e^{\alpha x}+b~.
\end{eqnarray}

To obtain the potentials $U_N(x)$ which admit solvable states with
higher weights $j$, we must first derive the Bethe ansatz
equations. To this end, let us perform the change of variable
$z=h(x)=\exp(-\alpha x)$. Eq.~(\ref{phi}) then becomes
\begin{eqnarray}
\left\{-\alpha z^2 \frac{d^2}{dz^2} +\left[2az^2 -(2b+\alpha)z
-2c\right]\frac{d}{dz} +\left[-2aNz -
\frac{\epsilon}{\alpha}\right]\right\}\phi(z) =0~. \label{phi-I}
\end{eqnarray}
For $N>0$, there are $N+1$ solutions which include excited states.
Assuming $\phi(z)=\prod_{i=1}^N (z-z_i)$ in Eq.~(\ref{phi-I}), one
obtains the Bethe ansatz equations which determine the roots
$z_i$'s
\begin{eqnarray}
2az_i^2 -(2b+\alpha)z_i - 2c -2\alpha \sum_{l\neq
i}\frac{z_i^2}{z_i-z_l} =0~, \quad\quad i=1,\ldots,N~,
\label{BA-I}
\end{eqnarray}
and the equation which gives the energy parameter in terms of the
roots $z_i$'s
\begin{eqnarray}
\epsilon =2\alpha c\sum_{i=1}^N \frac{1}{z_i} ~. \label{E-I}
\end{eqnarray}
Each set of \{$z_i$\} determines a QES energy $E$ with the
corresponding polynomial $\phi$.

As an example, let us construct $U_1(x)$ which admits two
solutions.  This corresponds to the case with $N=1$ and
$\phi(z)=z-z_1$. According to Eq.~(\ref{BA-I}), the root $z_1$
satisfies
\begin{eqnarray}
2az_1^2 -(2b+\alpha)z_1 -2c =0~,
\end{eqnarray}
which gives two solutions
\begin{eqnarray}
z_1^\pm=\frac{(2b+\alpha)\pm\sqrt{(2b+\alpha)^2 + 16ac}}{4a}~.
\end{eqnarray}
The corresponding energy parameters are
\begin{eqnarray}
\epsilon^\pm&=&  2\alpha c\frac{1}{z_1}\cr &=&
-\frac{\alpha}{2}\left[(2b+\alpha)\mp\sqrt{(2b+\alpha)^2 +
16ac}\right]~.
\end{eqnarray}
For the parameters assumed here, the solution with root
$z_1^-=-|z_1^-|<0$ gives the ground state, while that with root
$z_1^+>0$ gives the first excited state. The superpotential is
constructed according to Eq.~(\ref{W}) as
\begin{eqnarray}
U_1(x)&=&U_0-\frac{h^\prime(x)}{h(x)-z_1^-}\cr &=&-ae^{-\alpha x}+
c e^{\alpha x}+\frac{\alpha}{1+|z_1^-|e^{\alpha x}}+b~.
\end{eqnarray}
The ground state and the excited state have energy $
E^2=\epsilon^--\epsilon^-=0$ and $E^2=\epsilon^+-\epsilon^-=
\alpha \sqrt{(2b+\alpha)^2 + 16ac}$, respectively.

Potential $U_N(x)$ admitting $N+1$ solvable states can be
constructed accordingly.

\section{Three- and four-dimensional Dirac equations in Cartesian coordinates}

\subsection{Three-dimensional Dirac equation}

A $(2+1)$-dimensional Dirac Hamiltonian has the form
\begin{equation}
H=\balpha \cdot {\bf p} + \beta (m+V_s(x))~, \label{H2}
\end{equation}
where ${\bf p}=(p_x,p_y)=-i(d/dx, d/dy)$, and $\balpha,~\beta$ are
Dirac matrices.  The potential $V_s(x)$ is assumed to depend only
on $x$.

The Dirac equation can also be cast into factorized, or
supersymmetric form.  We represent the Dirac matrices in terms of
the Pauli matrices $\bsigma$ as follows:
\be
\alpha_x=\sigma_x~, ~~~\alpha_y=\sigma_y~,~~\beta=\sigma_z~. \ee
As $V(x)$ depends on $x$ only, the wave function can be written as
\be
\psi = e^{ik_yy} \left(\begin{array}{c} f_-(x)\\ f_+(x)
\end{array}\right)~,
\label{psi}
\ee
 where $k_y$ is a real constant, and $f_\pm$ are real functions of
 $x$.
 The Dirac equation becomes
\be
\left( \begin{array}{cc} U(x) & p_x-ik_y\\ p_x + ik_y & -U(x)
\end{array}\right)~\left(\begin{array}{c} f_-(x)\\ f_+(x)
\end{array}\right)=E\left(\begin{array}{c} f_-(x)\\ f_+(x)
\end{array}\right)~.
\label{Dirac2} \ee Here $U(x)$ is again defined as
$U(x)=V_s(x)+m$.
 To cast Eq.~(\ref{Dirac2}) into supersymmetric form, we
 transform the wave function by a unitary matrix $T$:
\be
\left(\begin{array}{c} f_-(x)\\ f_+(x)
\end{array}\right)\to \left(\begin{array}{c} i\psi_-(x)\\ \psi_+(x)
\end{array}\right)\equiv T^\dagger~\left(\begin{array}{c} f_-(x)\\ f_+(x)
\end{array}\right)~,
~~~T= \frac{1}{\sqrt{2}}\left( \begin{array}{cc} 1 & i\\ i & 1
\end{array}\right)~.
\ee
 Then the Hamiltonian transforms as
\be
H\to T^\dagger H T= \left( \begin{array}{cc} k_y &
i\left(-\frac{d}{dx}+U(x)\right)\\
-i\left(-\frac{d}{dx}+U(x)\right) & -k_y
\end{array}\right)~,
\ee
 and the Dirac equation becomes
\be
\left(\frac{d}{dx} + U(x)\right)\psi_- &=& \left(E+ k_y\right)
\psi_+~,\\ \left(-\frac{d}{dx} +  U(x)\right)\psi_+ &=& \left(E-
k_y\right) \psi_-~, \label{susy-2} \ee which is exactly in the
same form as Eq.~(\ref{susy}).  Hence the exact/quasi-exact
solvability of Eq.~(\ref{susy-2}) can be discussed as in the
one-dimensional case.

\subsection{Four-dimensional Dirac equation}

Let us now consider exact/quasi-exact solvability of
four-dimensional Dirac equation with a Lorentz scalar potential
dependent only on one spatial variable, say $z$.  Such system is
useful in describing $z$-dependent valence- and conduction-band
edge of semiconductors near certain points in the Brillouin zone
\cite{Junker}.

 The Hamiltonian has the
form
\be
H&=&\balpha \cdot {\bf p} + \beta (m+V_s(z))\nonumber\\ &=&\left(
\begin{array}{cc} 0 & \balpha\cdot{\bf p}-i U(x)\\ \balpha\cdot{\bf p}+i U(x) &
0
\end{array}\right)~,
 \label{H3}
\ee where ${\bf p}=(p_x,p_y,p_z)=-i(d/dx,~d/dy,~d/dz)$. Here
$U(z)=V_s(z)+m$ as before, and we have chosen the Dirac matrices
in the supersymmetric representation \cite{Junker}
\be
{\balpha} = \left( \begin{array}{cc} 0 & {\bsigma}\\ {\bsigma} & 0
\end{array}\right)~,~~~~~
\beta= \left( \begin{array}{cc} 0 & -i \\ i & 0
\end{array}\right)~.
\ee

To see that the Hamiltonian is factorizable, it is more convenient
to consider its square $H^2$:
\be
H^2= \left( \begin{array}{cc} H^- & 0\\ 0 & H^+
\end{array}\right)~,
\ee
 where
\be
H^\mp\equiv -\nabla^2 + U^2(z) \pm U^\prime (z) \sigma_z ~.
\ee
 Let the wave function be in the form
\be
\psi = e^{ik_x x + ik_yy} \left(\begin{array}{c} \psi_-(z)\chi_-\\
\psi_+(z)\chi_+
\end{array}\right)~.
\ee
  Here  $k_x$ and $k_y$ are real constants,
$\psi_\pm (z)$ are real functions of $z$,  and $\chi_\pm$ are
two-component eigen-spinors of $\sigma_z$ :
$\sigma_z\chi_\pm=\pm\chi_\pm$.  Then the eigenvalues problem of
$H^2\psi=E^2\psi$ reads
\be
\left(-\frac{d^2}{dz^2}+U^2 \mp U^\prime\right)~\psi_\mp =
\epsilon \psi_\mp~, \label{susy-3} \ee
  with $\epsilon=E^2-k_x^2 - k_y^2$.  Now it is obvious that
Eq.~(\ref{susy-3}) is in the supersymmetric form Eq.~(\ref{susy}),
and construction of QES potentials $U(z)$ proceeds as in the
previous cases.

\section{Dirac equation with a Lorentz scalar and a magnetic monopole potential}

Dirac equation with a spherical Lorentz scalar potential was
originally motivated by the MIT bag model intended for describing
quark confinement \cite{MIT-bag,Bhad}.  Now we would like to
determine the forms of spherical Lorentz scalar potentials which
admit exact/quasi-exact solutions.

Generally, in the presence of an electromagnetic field $(V,{\bf
A})$ and a Lorentz scalar potential $V_s$, the Dirac equation is
given by
\be
H&=&\balpha \cdot \left({\bf p}- ie{\bf A}\right) + \beta U + eV~,
 \label{H4}
\ee
 where $e$ is the charge of the fermion and $U\equiv m+V_s$.
Quasi-exact solvability of the system when $V=V_s=0$ has been
demonstrated in connection with the Pauli equation \cite{HoRoy1}.
In the absence of the vector potential $\bf A$ and $V$, the system
has been shown in the last two sections to be QES in one spatial
dimension. In other coordinates, it is rather difficult, if not
impossible, to find QES potentials.  In this section, we shall
consider the system in spherical coordinates.

When ${\bf A}=0$, the Dirac equation in spherical coordinates can
be separated into a two-component radial equation \cite{Gross}
\be
\left( \begin{array}{cc} \frac{d}{dr}+\frac{\kappa}{r} & eV-E-U\\
eV-E +U & -\frac{d}{dr}+\frac{\kappa}{r}
\end{array}\right)~\left(\begin{array}{c} f(r)\\g(r)
\end{array}\right)=0~.
\label{Dirac3}
\ee
 The $\kappa$ in the centrifugal term is related
to the total angular momentum quantum number $j=1/2,~3/2,\ldots$
as $\kappa=\pm (J+1/2)$. It follows that $\kappa$ is nonzero
integer: $\kappa=\pm 1,~\pm 2,\ldots$.  As mentioned in the
Introduction, QES potentials with both $V$ and $U$ are difficult
to find.  The first example was presented in \cite{BK}, in which
$V$ is taken to be Coulomb-like and $V_s$ is linear in $r$.
Existence of infinitely many QES $V$ and $V_s$ with two solutions
based on some reparametrization scheme was shown in \cite{Znojil},
and explicit construction of a subclass of such doublet (and
multiplet) QES potentials were presented in \cite{SH}.

We want to determine possible forms of exactly solvable and QES
potentials based on factorizability of the system.  Unfortunately,
the Dirac equation Eq.~(\ref{Dirac3}) is difficult to factorize,
owing to the presence of the non-vanishing centrifugal term
$\kappa/r$ \cite{Nogami}. Nevertheless, it has been shown in
\cite{Torres} that $\kappa$ could be zero in the presence of a
Dirac magnetic monopole described by
\be
A_r=A_\theta=0~,~~~A_\phi=g\frac{1-\cos\theta}{r\sin\theta}~. \ee
Here $g$ is the magnetic charge of the Dirac monopole. In
\cite{Torres} it is shown that
\be
j = \left\{ \begin{array}{ll}
               \frac{1}{2}, ~ \frac{3}{2},\ldots &\mbox{if}~ q=0~;\\
               |q|-\frac{1}{2}~,~|q|+\frac{1}{2}~,~|q|-\frac{3}{2}~,\dots
                & \mbox{if }~q\neq 0~,
               \end{array}
               \right.
\ee
 where $q\equiv eg = 0,~\pm 1/2,~1,\ldots$ according to Dirac's
 quantization condition, and
\be
 \kappa = \pm \sqrt{\left(j + \frac{1}{2}\right)^2 - q^2}~.
 \label{kappa}
\ee
 Hence, $\kappa=0$ when $j=|q| - 1/2$, which is possible only
when the magnetic charge $g\neq 0$.  In this case, the Dirac
equation with a scalar and a Lorentz scalar potential in the
presence of the Dirac monopole field is also reducible to the
two-component radial equation (\ref{Dirac3}), with $\kappa=0$ (see
\cite{Torres} for technical details).

We now show that Eq.~(\ref{Dirac3}) is factorizable when $V=0$ in
the sector $j=|q|-1/2$ ($\kappa=0$).  In fact, if the wave
function is transformed to
\be
\left(\begin{array}{c} f\\g
\end{array}\right)\to \left(\begin{array}{c}\psi_-\\\psi_+
\end{array}\right)\equiv T^\dagger ~\left(\begin{array}{c} f\\g
\end{array}\right)~,~~~T= \frac{1}{\sqrt{2}}\left( \begin{array}{cc} 1 & 1\\ -1 & 1
\end{array}\right)~,
\ee
 then the Dirac equation (\ref{Dirac3}) changes to the
 following factorized form:
\be
\left(\frac{d}{dr} +  U(r)\right)\psi_- &=& E\psi_+~,\nonumber\\
\left(-\frac{d}{dr} +  U(r)\right)\psi_+ &=& E\psi_-~.
\ee
 Here $U(r)$ plays the role of the superpotential $W(r)$, and the
 energy parameter is $\epsilon = E^2$.
 The form of $U(r)$ admitting exact/quasi-exact solutions can then
 be constructed according to the prescribed procedure.

\subsection{Exactly solvable cases}

From the Table~4.1 of \cite{Cooper}, it is found that there are
four types of exactly solvable potentials $U(r)$ for this system,
namely, (i) 3D oscillator-like ; (ii) Coulomb-like; (iii) Eckart
potential; and (iv) generalized P\"oschl-Teller.

\subsection{Quasi-exactly solvable cases}

Three types of QES potentials $U(r)$ can be constructed, namely,
Class VII, VIII, and IX in \cite{Tur}.  We outline the
construction of Class VII below.

The general potential in Class VII has the form
\be
V_N(r)=a^2r^6+2abr^4+\left[b^2-a\left(4N+2\gamma
+3\right)\right]r^2
+\gamma\left(\gamma-1\right)r^{-2}-b\left(2\gamma+1\right)~,
\label{cVII} \ee where $a,b$ and $\gamma$ are constants. The gauge
function is
\be
g(r)=\frac{a}{4}r^4 + \frac{b}{2}r^2 -\gamma\ln {r}~. \label{g4}
\ee We must have $a,\gamma >0$ to ensure normalizability of the
wave function. The potential
\be
U_0(r)= g^\prime(r)= ar^3 + b r -\frac{\gamma}{r} \ee
 admits
a QES ground state with energy $E=0$, and the corresponding wave
function with components $\psi_-\propto \exp(-g_0(r))$ and
$\psi_+=0$.

To determine potentials admitting QES potentials $V_N$ with higher
weights, we need to obtain the Bethe ansatz equations for $\phi$.
Letting $z=h(r)=r^2$, Eq.~(\ref{phi}) becomes
\begin{eqnarray}
\left[-4 z \frac{d^2}{dz^2} +\left(4az^2 +4bz -2\left(2\gamma
+1\right) \right)\frac{d}{dz} -\left(4aNz + \epsilon
\right)\right]\phi(z) =0~. \label{phi-VII}
\end{eqnarray}
For $N=0$, the value of the $\epsilon$ is $\epsilon=0$. For higher
$N>0$ and $\phi(r)=\prod_{i=1}^N (z-z_i)$, the potential $U_N(r)$
is obtained from Eqs.~(\ref{W}):
\begin{eqnarray}
U_N(r) = U_0(r) - \sum_{i=1}^N \frac{h^\prime(r)}{h(r)-z_i}~.
\label{EN}
\end{eqnarray}
For the present case, the roots $z_i$'s are found from the Bethe
ansatz equations
\begin{eqnarray}
2az_i^2 +2bz_i -\left(2\gamma+1\right) - \sum_{l\neq
i}\frac{z_i}{z_i-z_l} =0~, \quad\quad i=1,\ldots,N~,
\label{BA-VII}
\end{eqnarray}
and $\epsilon$ in terms of the roots $z_i$'s is
\begin{eqnarray}
\epsilon=2\left(2\gamma+1\right)\sum_{i=1}^N \frac{1}{z_i}~.
\label{E-VII}
\end{eqnarray}

For $N=1$ the roots $z_1$ are
\begin{eqnarray}
z_1^\pm=\frac{-b\pm\sqrt{b^2+2a(2\gamma+1)}}{2a}~, \label{z1}
\end{eqnarray}
and the values of $\epsilon$ are
\begin{eqnarray}
\epsilon^\pm=2\left(b\pm\sqrt{b^2+2a(2\gamma+1)}\right)~.
\end{eqnarray}
 For $a>0$, the root $z_1^-=-|z_1^-|<0$ gives the
ground state. With this root, one gets the potential
\begin{eqnarray}
U_1(r)=ar^3 +br -\frac{2r}{r^2+|z_1^-|} -\frac{\gamma}{r}~.
\end{eqnarray}

QES potentials  for higher degree $N$ can be constructed in the
same manner.

\section{Summary}

The exact/quasi-exact solvability of the Dirac equation with a
Lorentz scalar potential is examined in this paper.  Possible
forms of the potentials allowing factorization of the Dirac
equation are identified in two, three, and four dimensions.  Based
on such factorizability, we have classified all exactly solvable
potentials, and QES potentials based on $sl(2)$ Lie-algebra
according to Turbiner's calssification.

In this work and in \cite{HoRoy1,HoRoy2}, we have classified
exactly solvable and QES potentials in the Pauli equation and the
Dirac equation coupled minimally to a stationary vector potential,
the Dirac-Pauli equation (equivalent to generalized Dirac
oscillator) coupled non-minimally to an external electric fields,
and Dirac equation with a Lorentz scalar potential. Nonetheless,
we should mention that we have by no means exhausted all
possibilities of exact/quasi-exact potentials. Our classification
and construction are based on the factorizability, or
supersymmetric structure, of the system, with $sl(2)$ as the
underlying symmetry. There are QES systems which are not
factorizable, such as those considered in \cite{BK,BN,Znojil,SH},
and systems which are not related to the Lie-algebra $sl(2)$, such
as that discussed in \cite{Ho}.  It is an interesting and
challenging task to develop new methods to classify and construct
QES potentials in multi-component wave equations with any
Lie-algebraic structures.

\begin{acknowledgments}
This work was supported in part by the Republic of China through
Grant No. NSC 94-2112-M-032-007.
\end{acknowledgments}

\end{document}